\documentclass[%
 reprint,
superscriptaddress,
 amsmath,amssymb,
 aps,
prb,
]{revtex4-2}

\usepackage{graphicx}
\usepackage{dcolumn}
\usepackage{bm}
\usepackage{xcolor}
\usepackage{hyperref}

\begin{document}

\preprint{APS/123-QED}

\title{Coupled sawtooth chain exchange network in olivine Mn$_2$GeO$_4$}

\author{Vincent C. Morano}
\email{vincent.morano@psi.ch}
\affiliation{PSI Center for Neutron and Muon Sciences, Forschungsstrasse 111, 5232 Villigen, PSI, Switzerland}
\author{Zeno Maesen}
\affiliation{PSI Center for Neutron and Muon Sciences, Forschungsstrasse 111, 5232 Villigen, PSI, Switzerland}
\author{Stanislav Nikitin}
\affiliation{PSI Center for Neutron and Muon Sciences, Forschungsstrasse 111, 5232 Villigen, PSI, Switzerland}
\author{Jonathan S. White}
\affiliation{PSI Center for Neutron and Muon Sciences, Forschungsstrasse 111, 5232 Villigen, PSI, Switzerland}
\author{Takashi Honda}
\affiliation{Institute of Materials Structure Science, High Energy Accelerator Research Organization (KEK), Tokai, Ibaraki 319-1106, Japan}
\author{Tsuyoshi Kimura}
\affiliation{Department of Applied Physics, University of Tokyo, Bunkyo-ku, Tokyo 113-8656, Japan}
\author{Michel Kenzelmann}
\affiliation{PSI Center for Neutron and Muon Sciences, Forschungsstrasse 111, 5232 Villigen, PSI, Switzerland}
\author{Daniel Pajerowski}
\affiliation{Spallation Neutron Source, Oak Ridge National Laboratory, Tennessee, USA}
\author{Oksana Zaharko}%
\affiliation{PSI Center for Neutron and Muon Sciences, Forschungsstrasse 111, 5232 Villigen, PSI, Switzerland}

\date{\today}

\begin{abstract}
Sawtooth chain magnets have been a subject of historical interest in the field of frustrated magnetism, with classical olivine family $M_2TX_4$, ($M$ - 3d, $T$ - 4p, $X$ - chalcogen elements) typically realizing simple $\mathbf{k} = (000)$ states. The magnetism of the Mn$_2$GeO$_4$ olivine is surprisingly complex, proceeding from commensurate states to a multiferroic commensurate + incommensurate phase. Here we report inelastic neutron scattering results from a Mn$_2$GeO$_4$ single crystal and develop an effective Hamiltonian including long-distance bilinear and dipolar interactions. The magnetic interactions are predominantly antiferromagnetic and span a three-dimensional exchange network consisting of coupled sawtooth chains. Based on the determined strength of the couplings, the dominant sawtooth chains appear at third- and fourth- rather than next-nearest-neighbor. However the next-nearest-neighbor interaction is, along with a modest Dzyaloshinskii-Moriya interaction, important for modeling the observed incommensurability. We use the best-fit Hamiltonian as the basis for Langevin dynamics simulations and Luttinger-Tisza calculations of the high-temperature commensurate transition.
\end{abstract}

\maketitle


\section{Introduction}

Geometrically frustrated magnets are defined in the limit of weak disorder and strong frustration \cite{Ramirez1994}. While the cooperative magnetic physics emerging from their competing interactions is distinct from that of traditional ferromagnets (FM), antiferromagnets (AFM), and ferrimagnets, it remains an intrinsic consequence of the geometry of the lattice rather than the result of disorder. Triangular motifs are ubiquitous in such lattices, with triangular, kagome, face-centered cubic, and pyrochlore lattices potentially giving strong geometrical frustration. To identify other lattices that are suitable for producing geometrical frustration in real materials, it can be insightful to consider naturally occurring minerals.

\begin{figure*}
\includegraphics[width=0.99\textwidth]{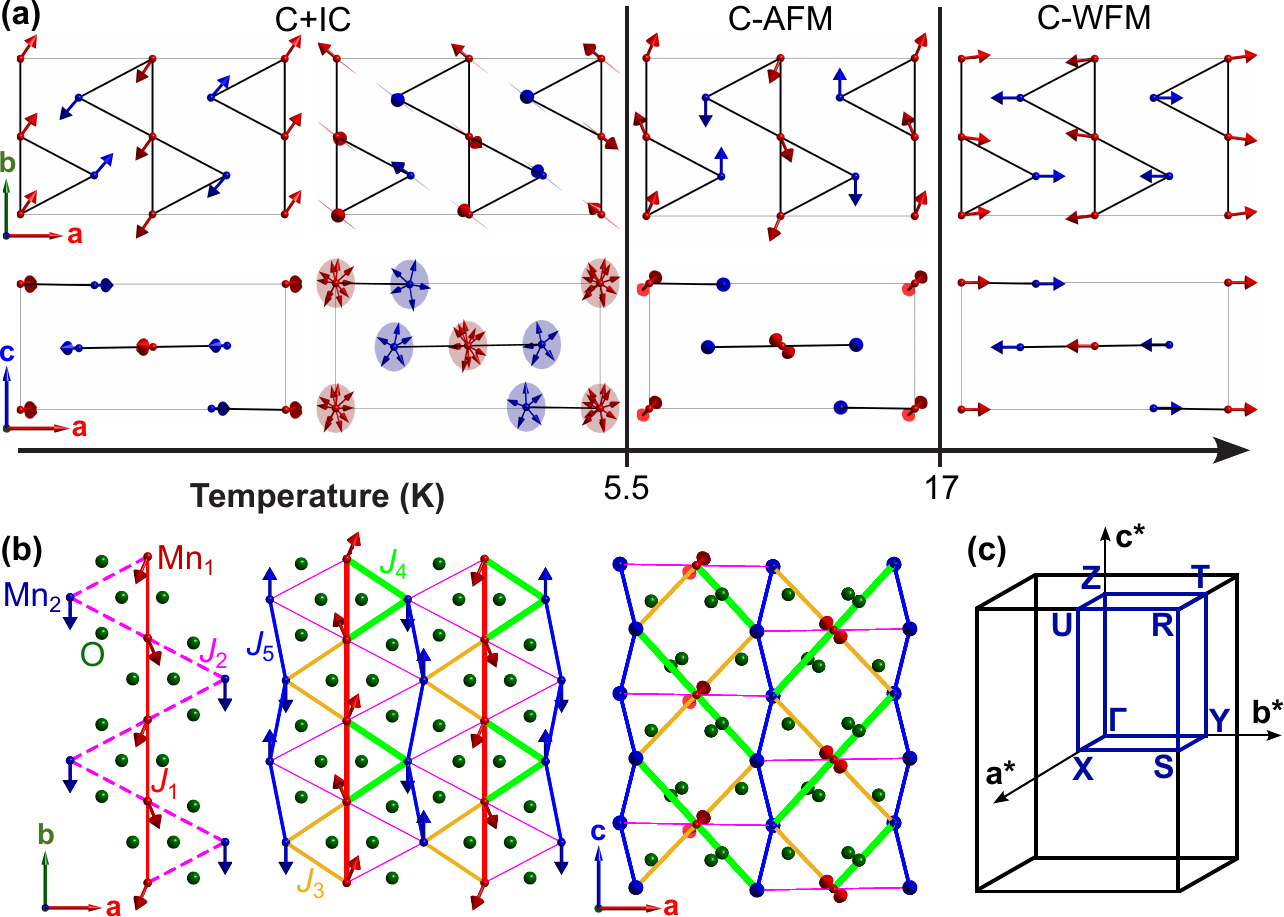}
\caption{\label{fig:Phases}
(a) Below $T_{\textrm{N1}} = 47$ K, Mn$_2$GeO$_4$ orders into a collinear $\mathbf{k} = (000)$ state with the moments primarily along $\mathbf{a}$. At $T_{\textrm{N2}} = 17$ K, the moments reorient to point primarily along $\mathbf{b}$. Finally at $T_{\textrm{N3}} = 5.5$ K, an additional $\mathbf{k}_2 = (0.136,0.211,0)$ IC propagation vector is observed while the $\mathbf{k}_1 = (000)$ component gains contributions from both the C-AFM and C-WFM irreps. The unit cell and sawtooth chains extending along $\mathbf{b}$ are labeled with black lines. The panel depicting the IC $ac$-plane extends across five unit cells along the $\mathbf{b}$-axis. The circles are a guide-to-the-eye for the moments' rotation plane. (b) The left image shows the nearest-neighbor $J_1$ and next nearest-neighbor $J_2$ interactions form sawtooth chains extending in one dimension along $\mathbf{b}$. $J_1$ couples Mn$_1$ sites while $J_2$ couples Mn$_1$ and Mn$_2$ sites. The middle and right images demonstrate that these chains are linked by more distant interchain superexchange interactions $J_{3-5}$ forming a three-dimensional exchange network with components along $\mathbf{c}$. When looking along $\mathbf{b}$, it becomes apparent that the three sawtooth chains ($J_1-J_2$ in magenta, $J_1-J_3$ in orange, and $J_1-J_4$ in green) run collinearly about the same Mn$_1$ (red) chains spanned by $J_1$. The ``teeth" of these chains also meet at Mn$_2$ (blue) to form sawtooth ``planes" in $bc$. At $T = 10$ K, the staggered component of the Mn$_1$ moment when viewed along the $\mathbf{b}$-axis lies along $J_3$ but perpendicular to $J_4$. Here we show the magnetic structure of the C-AFM phase. The oxygen atoms are displayed in green, while germanium atoms are not shown. The thickness of the bonds in the middle and right structures is proportional to the strengths reported in Table \ref{tab:interactions}. (c) The Brillouin zone for orthorhombic space group 62 ($Pnma$).}
\end{figure*}

Olivines, with chemical formula $M_2TX_4$, ($M$ - 3d, $T$ - 4p, $X$ - chalcogen elements), are the most prevalent mineral in Earth's mantle, and are common in both the Moon and stony meteorites \cite{Elias2020}. While the most abundant naturally occurring olivines span compositions between Mg$_2$SiO$_4$ (forsterite) and Fe$_2$SiO$_4$ (fayalite), the olivine-group includes isostructural compounds that can be either naturally-occurring (e.g. Mn$_2$SiO$_4$, tephroite) or synthetic (e.g. Mn$_2$GeO$_4$). Olivines have been rather well-characterized because of their importance to the structure of the Earth. Their crystal structure is orthorhombic, belonging to space group $Pnma$ (62). The crystal unit cell (Fig. \ref{fig:Phases}a) contains two distinct Wyckoff sites for the transition metal ions: $T_1$ (4$a$) at an inversion center and $T_2$ (4$c$) at a mirror plane. The nearest-neighbor (nn) interaction $J_1$ connects the $T_1$ atoms along the $\mathbf{b}$-axis, while the next nearest-neighbor (nnn) bond $J_2$ connects $T_1$ and $T_2$ atoms (Fig. \ref{fig:Phases}b). Together, these two bonds form sawtooth chains extending along $\mathbf{b}$ \cite{Hagemann2000, Lau2006}. The realization that such chains of corner-sharing isosceles triangles, which can be understood as a partial kagome lattice, could give geometrical frustration motivated further exploration of the olivines. In this work, we focus on the synthetic olivine Mn$_2$GeO$_4$.

Although subsequent magnetic characterization demonstrated geometrical frustration is weak in Mn$_2$GeO$_4$ (frustration factor $f \equiv |\Theta_\mathrm{CW}|/T_\mathrm{N} = 3.4$ \cite{Honda2012}), the magnetic phase diagram was found to be rich with three magnetic phase transitions upon cooling \cite{White2012}.

1) At $T_\mathrm{N1} = 47$ K, the magnetic Mn$^{2+}$ ions ($S=5/2, L=0$) order with the magnetic unit cell equal to the crystal one, $\mathbf{k}=(000)$, and the moments nearly collinear along the $\mathbf{a}$-axis. Each $J_1-J_2$ sawtooth chain is nearly FM, with a small staggered magnetization along $\mathbf{b}$. There is also weak ferromagnetism along $\mathbf{c}$, we therefore refer to this as the ``C-WFM" phase in accordance with reference \cite{Honda2012}. This order is consistent with the magnetic space group $Pn'm'a$ (62.446) in the BNS setting \cite{ISOMAG} where weak ferromagnetism is allowed.

2) At $T_\mathrm{N2} = 17$ K, the moments reorient to a purely AFM $\mathbf{k}=(000)$ configuration, ``C-AFM". While to a rough approximation the reorientation involves a $90^\circ$ rotation, the structure also becomes increasingly noncollinear with the Mn$_1$ sites acquiring a significant staggered magnetization along both $\mathbf{a}$ and $\mathbf{c}$. This arrangement is described by the magnetic space group $Pnma$ (62.441), for which no FM component is allowed.

3) Finally below $T_\mathrm{N3} = 5.5$ K, Mn$_2$GeO$_4$ realizes a $\mathbf{k}_1 = (000)$, $\mathbf{k}_2 = (0.136,0.211,0)$ transverse conical commensurate (C) + incommensurate (IC) state \cite{Harris2017}, ``C+IC". Here the C component of the magnetism is described by a combination of the irreducible representations from the high-temperature C-WFM and C-AFM phases such that a FM component is allowed and indeed reappears along $\mathbf{c}$. The IC component of the magnetism breaks inversion symmetry to give a ferroelectric (FE) polarization also along $\mathbf{c}$, realizing an uncommon form of multiferroicity in which both FE and FM polarizations are directed along the same axis. Although often described as a C+IC state, the C and IC components combine to produce a multi-$\mathbf{k}$ conical spiral, where the C part sets the cone axis and the IC part describes a spin spiral \cite{White2016, Honda2017}.

We study the Mn$_2$GeO$_4$ excitation spectrum by inelastic neutron scattering (INS) with the aim of understanding how this variety of magnetic phases emerges. We consider an effective Hamiltonian with Heisenberg interactions up to the tenth nn and find the first five (AFM) interactions dominate. Notably, the spectrum cannot be reproduced by isolated $J_1-J_2$ sawtooth chains \cite{Honda2012, Volkov2013}. Rather the exchange network is three-dimensional, consisting of coupled $J_1-J_3$ and $J_1-J_4$ sawtooth chains. We observe key features in the spectrum signaling the onset of C+IC order below $T = 5.5$ K. Despite the weak geometrical frustration, we identify competing interactions driving the incommensurability and concomitant onset of multiferroic order. Finally, using the best-fit Hamiltonian as the input for Langevin dynamics simulations \cite{Dahlbom2022, Dahlbom2025} and Luttinger-Tisza calculations \cite{Santoro1966}, we interpret the $T_\mathrm{N2}$ spin reorientation between C phases.

\section{Experiment}

Single crystals were grown by floating zone at Osaka University and characterized by X-ray diffraction, magnetization, dielectric constant, and electric polarization \cite{White2012}. The INS experiment was performed on the Cold Neutron Chopper Spectrometer (CNCS) at the Spallation Neutron Source in Oak Ridge National Laboratory \cite{Ehlers2011}. We aligned a $m = 0.47$ gram Mn$_2$GeO$_4$ crystal (studied in reference \cite{White2012}) in the $(0KL)$ horizontal scattering plane, and mounted it in a 5 T vertical field cryomagnet with a base temperature of $T = 1.8$ K. Here we focus on the 0-field results obtained with initial neutron energy $E_\textrm{i} = 12$ meV. Datasets spanning the Brillouin zone (Fig. \ref{fig:Phases}c) were collected in the C-WFM phase at $T = 20$ K, in the C-AFM phase at $T = 10$ K, and in the C+IC phase at base temperature. Data reduction and analysis were performed using the Horace \cite{Ewings2016} and SpinW \cite{Toth2015} libraries in MATLAB, and the Sunny \cite{Dahlbom2025} library in Julia. All data have been folded into the positive $(HKL)$ octant appropriate for orthorhombic symmetry. The code used for analysis is publicly available at \cite{morano_2025_17791877}.

\begin{table}[b]
\caption{\label{tab:interactions}%
Bond lengths and strengths in the C-AFM phase. Positive values indicate AFM interactions. Note that the structure is given by the parameters obtained at $T = 7$ K in \cite{White2012}. ``Fit" indicates the best-fit exchange constants while ``AFM" indicates the purely antiferromagnetic model.}
\begin{ruledtabular}
\begin{tabular}{lccrr}
\textrm{Bond}&
\textrm{Distance (\AA)}&
\textrm{Atoms}&
\textrm{Fit (meV)}&
\textrm{AFM (meV)}\\
\colrule
1 & 3.14 & Mn$_1$-Mn$_1$ & 0.478(1) & 0.478\\
2 & 3.37 & Mn$_1$-Mn$_2$ & 0.085(2) & 0.085\\
3 & 3.77 & Mn$_1$-Mn$_2$ & 0.331(5) & 0.331\\
4 & 3.83 & Mn$_1$-Mn$_2$ & 0.584(3) & 0.584\\
5 & 4.08 & Mn$_2$-Mn$_2$ & 0.303(1) & 0.303\\
6 & 5.06 & Mn$_1$-Mn$_1$ & -0.034(1) & 0\\
7 & 5.06 & Mn$_2$-Mn$_2$ & -0.017(1) & 0\\
8 & 5.58 & Mn$_1$-Mn$_2$ & -0.082(2) & 0\\
9 & 5.68 & Mn$_2$-Mn$_2$ & 0.0808(9) & 0.0808\\
10 & 5.83 & Mn$_1$-Mn$_2$ & 0.031(1) & 0.031\\
\end{tabular}
\end{ruledtabular}
\end{table}

\section{Results}

\subsection{INS results}
\begin{figure*}
\includegraphics[width=0.99\textwidth]{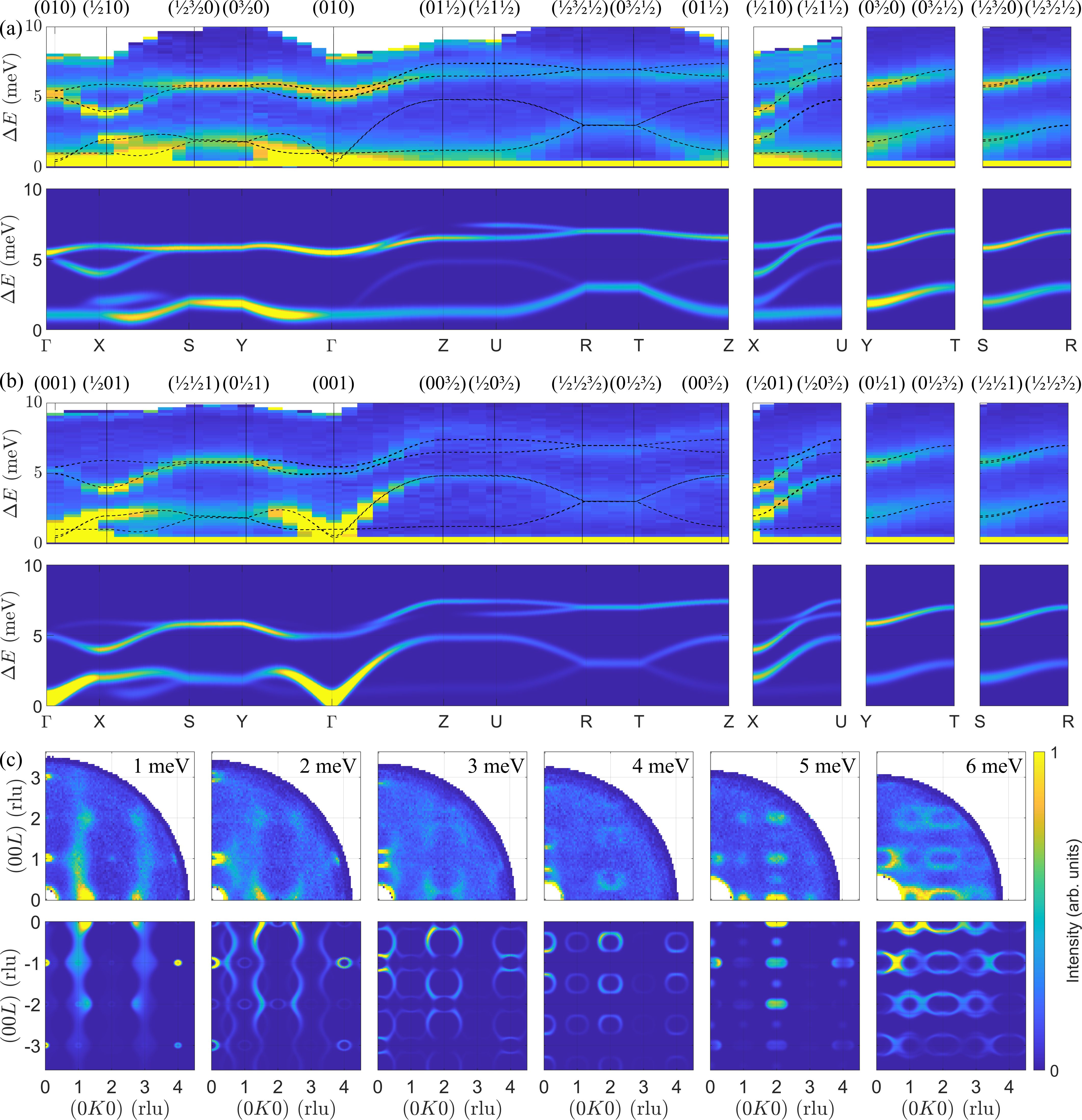}
\caption{\label{fig:10K} $T = 10$ K spectra (a) with $\Gamma$ taken to be (010) and (b) obtained with $\Gamma$ taken to be (001). Upper panels from (a,b) include the data obtained on CNCS while the lower panels are simulations from Eq. \ref{eq:ham}. The dispersion from these simulations is plotted atop the CNCS data as a dashed black line. Note the low-lying intensity along $\overline{\mathrm{X} \mathrm{S}}$ and $\overline{\mathrm{Y} \Gamma}$. We expect the intensity at the highest $\Delta E$'s, corresponding to the detector edges, are artifacts. (c) Constant-$\Delta E$ data (above) and simulations (below) from Eq. \ref{eq:ham}. The data have been corrected for the Bose population factor at $T = 10$ K.}
\end{figure*}

We begin by considering the excitation spectra of the C $\mathbf{k}=(000)$ phases. In the $T = 10$ K dataset collected below $\Delta E = 10$ meV (Fig. \ref{fig:10K}) four modes are observed, forming two distinct bands centered around $\Delta E = 2$ and 5 meV. Because there are eight magnetic ions in the magnetic unit cell, we expect each of the four modes to be doubly degenerate. There appears to be an additional degeneracy along the $\overline{\mathrm{SY}}$ and $\overline{\mathrm{RT}}$ paths. The modes disperse in all three dimensions, although there are some paths where the dispersion is small or nonexistent. From $\Delta E$-cuts at momentum transfer $\mathbf{Q} = (010)$ (see Supplemental Material Figure S2 and references therein \cite{Hardy1965, Deiseroth2005, Yamazaki1999, Momma2008}), we place an upper bound on the anisotropy gap $\Delta E \leq 0.2$ meV.

The $T = 20$ K dataset shown in Figure S4 is very similar. This suggests that the magnetic interactions do not change dramatically upon cooling through the $T_\textrm{N2} = 17$ K transition.

\begin{figure*}
\includegraphics[width=0.99\textwidth]{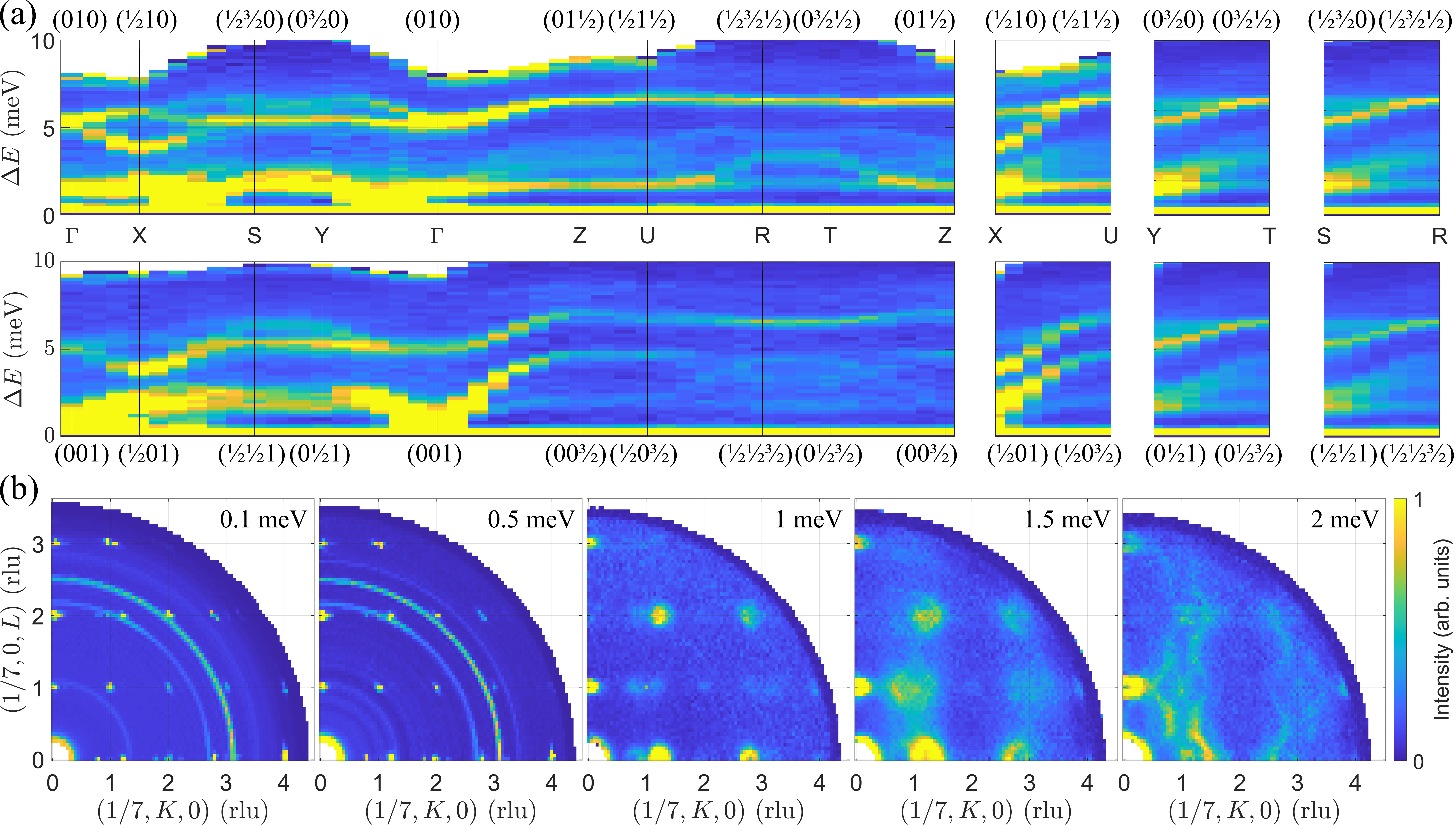}
\caption{\label{fig:2K}(a) $T = 2$ K spectra with $\Gamma$ taken to be (010) in the upper panel and (001) in the lower panel. We expect the intensity at the highest $\Delta E$'s, corresponding to the detector edges, is an artifact. (b) Constant-$\Delta E$ slices showing coherent spin wave excitations emerging at finite energy transfer above the IC $\mathbf{k}_2 = (0.136,0.211,0)$ Bragg peaks at the elastic line. Compare to the finite $\Delta E$ modes at $T = 10$ K in Figure \ref{fig:10K}(c) extending to the IC positions. There are powder rings near the elastic line and artifacts at the low-$\mathbf{Q}$ detector edges. The data have been corrected for the Bose population factor at $T = 2$ K.}
\end{figure*}

At $T = 2$ K (Fig. \ref{fig:2K}), in addition to the modes located at the integer-valued momentum transfers, further excitations arise from the IC positions $\mathbf{k}_2 = (0.136,0.211,0)$ (Fig. \ref{fig:2K}b). They correspond to the multi-$\mathbf{k}$ state emerging below $T_\mathrm{N3} = 5.5$ K \cite{White2012}.

\subsection{Exchange Hamiltonian modeling}

A simple Heisenberg $J_1$-$J_2$ model with sawtooth chains extending along $\mathbf{b}$ produces modes with dispersion only along the chain direction $\mathbf{b}$, i.e. $\mathbf{b}^* \propto (0K0)$ in reciprocal space. This is in clear contrast to the observed spectra with dispersion along $(H00)$, $(0K0)$ and $(00L)$, thus the $J_{1,2}$ model is not valid. The anisotropy of the $S = 5/2, L = 0$ Mn$^{2+}$ ions should be weak and primarily result from dipole-dipole contributions rather than single-ion anisotropy \cite{Keffer1952}. Therefore, we extend the Heisenberg model to further long-distance bilinear and dipolar interactions by considering the following Hamiltonian:

\begin{equation} \label{eq:ham}
    \mathcal{H}_0 = \sum_{i<j} J_{ij} \mathbf{S}_i \cdot \mathbf{S}_j - \frac{\mu_0}{4 \pi} \sum_{i<j} [3 (\boldsymbol{\mu}_i \cdot \hat{\mathbf{r}}_{ij}) (\boldsymbol{\mu}_j \cdot \hat{\mathbf{r}}_{ij}) - \boldsymbol{\mu}_i \cdot \boldsymbol{\mu}_j]/r_{ij}^3.
\end{equation}

\noindent
$\boldsymbol{\mu}_i$ is the magnetic moment on atom $i$, $\mathbf{r}_{ij}$ is the displacement between atoms $i$ and $j$, and $\mu_0$ is the permeability of free space. The fitted exchange constants up to the tenth nn interaction are reported in Table \ref{tab:interactions} while the dipole-dipole anisotropy involves no fitted parameters. Further details on the fitting procedure are available in the Supplemental Material, see Figure S9 for the $J_{1-5}$ superexchange pathways.

For this $3d^5$ system, the Goodenough-Kanamori rules \cite{Goodenough1963} predict AFM rather than FM superexchange interactions. Indeed, the dominant fitted exchange interactions are AFM \cite{Volkov2013}. We find that $J_{1,3-5}$ are of similar magnitude, while $J_{2,6-10}$ are weak. The spectrum is not well-described in terms of isolated strongly-interacting $J_1-J_2$ sawtooth chains, rather the superexchange network consists of $J_1-J_3$ and $J_1-J_4$ sawtooth chains extending along $\mathbf{b}$. These sawtooth chains meet at the Mn$_2$ sites to span the $bc$-plane forming dual sawtooth ``planes" consisting of both corner- and edge-sharing triangles. Similarly, the strong $J_5$ exchange network can be visualized as a buckled rhombic net spanning Mn$_2$ sites across the $bc$-plane. The dual sawtooth planes are linked by this $J_5$ interaction (as well as by the weak $J_2$ interaction) along $\mathbf{a}$ to give fully three-dimensional magnetism consistent with the observed dispersion. Based on the influence of $J_1$ and $J_5$ on the spectrum, the lower-$\Delta E$ mode appears to primarily correspond to the Mn$_1$ sites while the upper-$\Delta E$ mode corresponds to the Mn$_2$ sites. Although much of the spectrum is captured by a $J_{1-5}$ model, the $J_9$ interaction is important for producing dispersion on the high-$\Delta E$ band when $L=1/2$. We briefly note that if the C-WFM structure is approximated as collinear, the system becomes altermagnetic with $J_{20}$, $J_{21}$ as the first pair of interactions driving a chiral magnon splitting. We find that such long-range interactions, however, are not required to model the inelastic spectrum for Mn$_2$GeO$_4$.

Although the fits reported above reproduce important features of the spectrum, the $T = 10$ K model Hamiltonian does not yield the experimentally identified magnetic structure when the spin configuration is first randomized and then relaxed. While the observed $\mathbf{k} = (000)$ propagation vector is stabilized and the relaxed structure is close to that observed in the experiment, simulations show that the Mn$_1$ moments are constrained to the $ab$-plane whereas in the experiment a component of the staggered magnetization along $\mathbf{c}$ is also observed. We find that a modest Dzyaloshinskii-Moriya interaction (DMI) (e.g. $\mathbf{D}_\mathrm{1a}$ = 0.01 meV, where here subscripts indicate the $\mathbf{a}$-component of the nn DMI) is able to generate the observed component along $\mathbf{c}$. The new Hamiltonian is given by

\begin{equation} \label{eq:hamDMI}
    \mathcal{H} = \mathcal{H}_0 + \sum_{\left<ij\right>} \mathbf{D}_{ij} \cdot \left( \mathbf{S}_i \times \mathbf{S}_j \right),
\end{equation}

\noindent
where $\left< ij \right>$ indicates a sum over nearest-neighbors. However, the inclusion of DMI on this scale does not have a significant impact on the simulated inelastic spectrum and therefore we presume that our experiment is not sensitive to this DMI component.

As for the incommensurability, we find that a modified Hamiltonian that is purely antiferromagnetic (see column ``AFM" in Table \ref{tab:interactions} and Figure S8a) does produce a minimum in the dispersion at low-$\Delta E$, with the $q_y$-component being close to the experimentally observed IC propagation vector $\mathbf{k}_2 = (0.136,0.211,0)$. Indeed this minimum is observed developing in the $T = 10$ K data (see the first panel of Figure \ref{fig:10K}c). While $J_2$ is weak we find that the $J_1-J_2$ sawtooth chains play an important role in stabilizing the incommensurability in $(0K0)$. For example, setting $J_2 = 0$ removes the minimum in $(0K0)$. Upon introducing a $\mathbf{b}$-component to the nn DMI, which becomes symmetry-allowed in the C+IC phase \cite{White2012}, we find that the minimum develops an additional nonzero component in $\mathbf{q} \approx (H, 0.2, 0)$ (Figure S8b). The position of this minimum in $H$ maybe be tuned by adjusting $\mathbf{D}_\mathrm{1b}$. Finally, static long-range C+IC order may be stabilized by modestly tuning the Heisenberg exchange parameters. For example, increasing $J_{1,2}$ from 0.478 meV and 0.08507 meV to 0.5 meV and 0.11 meV, respectively, condenses the low-lying spin wave into the elastic line at the IC position (Figure S8c).

In the following paragraphs, we use the Hamiltonian obtained by fitting the spin wave spectrum as a starting-point for modeling the $T_\mathrm{N2}$ spin reorientation.

\subsection{Langevin dynamics calculations}

We performed Langevin Dynamics (LD) calculations \cite{Dahlbom2022, Dahlbom2025} using the model Hamiltonian (Eq. \ref{eq:ham}) extracted at $T = 10$ K (column ``Fit" of Table \ref{tab:interactions}) with an Ewald summation for the dipole-dipole interactions to investigate what drives the reorientation of the moments observed at $T_\textrm{N2}$.

Recall that the transition at $T_\textrm{N2}$ can be roughly described as a 90$^\circ$ rotation of the moments from being principally along $\mathbf{a}$ to principally along $\mathbf{b}$. In the observed inelastic spectra, this reorientation does not appear to be coupled to a large change in the magnetic interactions. Could the weak dipole-dipole anisotropy, which is important for determining the overall orientation of the moments, cause the $T_\textrm{N2}$ transition?

LD simulations with the best-fit exchange parameters give a N\'eel transition with an onset of intensity at two selected magnetic reflections $(100)$ and $(120)$ (Fig. S5a). The presence of these reflections is consistent with the ground state observed when relaxing the structure at $T = 0$ K. However, there are no clear anomalies near $T_\textrm{N2}$ that would signal a phase transition upon cooling. We therefore tested a few deviations from the 10 K fit exchange parameters. First, as relatively small changes in exchange can significantly impact the magnetic order, we set the small fitted negative exchange constants to zero (column ``AFM" in Table \ref{tab:interactions}). In repeated LD runs, the calculated static structure factors for $(100)$ and $(120)$ show anomalies that qualitatively match the observed N\'eel $T_\mathrm{N1} = 47$ K and C $T_\mathrm{N2} = 17$ K phase transitions (Fig. S5b). Turning off the Ewald summation and rerunning the Langevin dynamics for the purely AFM exchanges, we find that the intermediate phase transition is no longer present (Fig. S5c). Based on the calculated structure factors, we hypothesize that dipole-dipole anisotropy is responsible for the spin reorientation at $T_\textrm{N2}$. The LD runs using the refined and modified $T = 10$ K Hamiltonian do not capture the $T_\mathrm{N3} = 5.5$ K transition which may be driven by DMI and/or a modification of the Hamiltonian upon cooling.

\section{Discussion}

Initially much of the motivation for studying magnetic olivines $M_2TX_4$, and Mn$_2$GeO$_4$ specifically, was the apparent geometrical frustration via the $J_1-J_2$ sawtooth chains \cite{Hagemann2000}. Yet, we find that while the $J_1$ interaction is strong and AFM, the $J_2$ interaction forming the ``teeth" of the $J_1-J_2$ sawtooth chains is only weakly AFM, despite the bonds sharing similar distances and angles (Fig. S9). To understand this difference one may need to account for the contribution of the $T$-atoms to the exchange paths, as was suggested for the olivine Cr$_2$BeO$_4$ \cite{Saji1974}. Apart from the $J_1-J_2$ sawtooth chains, triangular motifs are also formed by antiferromagnetic $J_1-J_3$ and $J_1-J_4$ sawtooth chains that contribute to the frustration (cf. classical sawtooth magnet Rb$_2$Fe$_2$O(AsO$_4$)$_2$ \cite{Garlea2014} and the quantum sawtooth magnet Cu$_2$Cl(OH)$_3$, atacamite \cite{Heinze2021, Heinze2025}).

This work provides microscopic insight into the interactions driving the C+IC state and the corresponding multiferroicity. In our model, the frustrated exchange network consists of sawtooth chains coupled in three dimensions. Although small, $J_2$ is essential for generating the incommensurate soft mode while DMI fixes its orientation. With the onset of static incommensurate order below $T_\mathrm{N3}$, both inversion and time-reversal symmetry are broken, giving rise to the observed multiferroicity. Indeed, other frustrated systems such as delafossites and spinels have been observed to adopt spiral (e.g. cycloidal) magnetic structures, with inversion and time-reversal symmetry breaking that yields a magnetoelectric multiferroic state \cite{Kimura2007}.

Our suggestion that the dipole-dipole anisotropy instigates the transition between the C states in  Mn$_2$GeO$_4$ is supported by other experimental observations. For example, powder neutron diffraction from the Supplemental Material of Ref.\cite{White2012} (Fig. S5d) shows the Mn$_2$ ordered moment saturates sooner than that of Mn$_1$, with each sublattice having a distinct $T$-dependence. Magnetization data \cite{Honda2012} show the phase boundary between the C-WFM and C-AFM states falling to $\mu_0 H = 0$ T at $T_\mathrm{N2}$ indicating the two phases are close in energy. The $T_\mathrm{N2}$ spin reorientation may therefore result from the change of the dipole-dipole anisotropy as the Mn$_1$ sublattice magnetization increases upon cooling.

Reference \cite{Santoro1966} develops an analytical theory for the collinear to noncollinear phase transition in the olivine family (reported at $T = 23$ for Fe$_2$SiO$_4$ \cite{Santoro1966, Tripoliti2023} and $T = 15$ K for Mn$_2$SiO$_4$ \cite{Santoro1966, Lottermoser1988}) based on the method of Luttinger and Tisza. This transition appears to be related to $T_\mathrm{N2}$ in Mn$_2$GeO$_4$. First, the transition temperatures in other olivines are rather close to $T_\mathrm{N2} = 17$ K. Second, the Mn$_2$GeO$_4$ C-WFM structure observed at high temperatures is nearly collinear while the C-AFM structure that develops upon cooling matches noncollinear olivine phases considered by \cite{Santoro1966}. The model of \cite{Santoro1966}, however, does not include dipole-dipole anisotropy. Even though the LD simulation without dipole-dipole anisotropy does not produce the observed static structure factors upon passing through $T_\mathrm{N2}$, could this analytical model capture some behavior of the $T_\mathrm{N2} = 17$ K transition in Mn$_2$GeO$_4$? Plugging in the best-fit $J_{1-5}$, we find the collinear magnetic structure is stable above a transition temperature $T_\mathrm{t}$ while the noncollinear structure is stable below. The transition temperature is determined by the relative magnetization of the two Mn sublattices $\rho = M_\textrm{I}/M_\textrm{II}$, offering a second mechanism by which the developing Mn$_1$ moment could trigger a spin reorientation. There are, however, certain caveats with the application of this model to Mn$_2$GeO$_4$. First, there is a small staggered magnetization observed in the C-AFM phase that makes the structure noncollinear, and a small FM component (as in Mn$_2$SiO$_4$ \cite{Lottermoser1988}). Second, the overall orientation of the moments (i.e. along $\mathbf{a}$ in the C-WFM phase) changes upon proceeding through the (approximately) collinear to noncollinear transition at $T_\mathrm{N2}$. Although both Mn$_2$SiO$_4$ and Fe$_2$SiO$_4$ are reported to undergo collinear to noncollinear spin reorientations, the C-WFM phase of Mn$_2$GeO$_4$ above $T_\mathrm{N2}$ closely resembles Mn$_2$SiO$_4$ \cite{Lottermoser1988} whereas the C-AFM structure resembles Fe$_2$SiO$_4$ \cite{Lottermoser1988, Tripoliti2023}. Indeed the magnetic space group symmetry of Mn$_2$GeO$_4$ changes from that of Mn$_2$SiO$_4$ (62.446) to that of Fe$_2$SiO$_4$ (62.441) when passing through $T_\mathrm{N2}$, in contrast to Fe$_2$SiO$_4$ which reportedly retains the same magnetic symmetry as it proceeds through the reorientation \cite{Tripoliti2023}. While the analytical model of \cite{Santoro1966} does captures important features of the transition at $T_\mathrm{N2}$, the inclusion of dipole-dipole anisotropy to the Hamiltonian may help resolve these peculiarities.

This work provides insight into the interactions that drive the unique physics of Mn$_2$GeO$_4$. Other olivines such as Mn$_2$Ge$X_4$ ($X$ = S, Se), $M_2$SiO$_4$ ($M$ = Fe, Co, Ni), Mn$_2$Si$X_4$ ($X$ = S, Se), and Cr$_2$BeO$_4$ present different magnetic structures and temperature-dependence \cite{Solzi2020, Deiseroth2005, Newnham1965, Colman2009, Kimura2012, Fujii2013, Nhalil2019, Mandujano2023, Mandujano2023b}. Additional inelastic work is therefore desirable both to comprehend their own underlying physics and to draw deeper comparisons with Mn$_2$GeO$_4$. Lithium orthophosphate LiNiPO$_4$ is also multiferroic and shares the same space group as the olivines $M_2TX_4$, but the magnetic $S = 1$ Ni$^{2+}$ ion occupies only the 4$c$ Wyckoff position corresponding to atom $M_2$ \cite{Vaknin2004, Jensen2009}. Diffuse scattering is observed up to about 2$T_\mathrm{N}$, with static long-range $\mathbf{k}_\mathrm{IC} = (0,0.12,0)$ order at $T_\mathrm{IC} = 21.7$ K replaced by C order at $T_\mathrm{N} = 20.8$ K. In both LiNiPO$_4$ and Mn$_2$GeO$_4$, the shallow minimum in the dispersion at the IC position is observed developing in the $\mathbf{k} = (000)$ phase. Competing interactions, here on the 4$c$ sublattice, also give the observed nonzero $K$-component of the incommensurability in LiNiPO$_4$. For LiNiPO$_4$, however, the incommensurate phase appears at high- rather than low-temperature and single-ion anisotropy is significant. Simulations in the FE phase of Mn$_2$GeO$_4$ will be valuable to develop quantitative agreement between the experimentally observed and simulated $\mathbf{k}$-vector components, and to understand the nature of the change in the $\mathbf{k} = (000)$ component of the magnetism below $T_\mathrm{N3}$ \cite{White2012}. Magnetoelastic effects \cite{Honda2012} and the origin of the difference in magnitude between $J_1$ and $J_2$ should be investigated more thoroughly \cite{Santoro1964, Saji1974}.

\section{Conclusion}

Here we have reported spin wave excitation data from an INS experiment on a Mn$_2$GeO$_4$ single crystal, and suggest a few effective Hamiltonians capturing various aspects of the magnetism. The INS experiment reveals a three-dimensional AFM exchange network of coupled sawtooth chains mediating the C spin reorientation and the onset of C+IC multiferroic order. Beyond Mn$_2$GeO$_4$ \cite{Honda2017, Leo2025} we expect our results will have broad applicability to the olivines, especially those for which Mn$^{2+}$ is the magnetic ion. This work suggests how frustration and DMI may lead to the emergence of spiral spin order and multiferroicity on sawtooth lattices that are not limited to the nn and nnn interactions. It contributes to the deeper understanding of magnetism on sawtooth chain lattices hosting strongly-coupled classical spins.

\begin{acknowledgments}
We acknowledge helpful discussions from Tom Fennell, Iurii Timrov, Flaviano Jos\'e dos Santos, and Randy S. Fishman. We gratefully acknowledge the contributions of Andrey Podlesnyak (CNCS), Louis-Pierre Regnault (IN22), Karin Schmalzl (IN12), and Christof Niedermayer (RITA-II) to previous unpublished INS experiments on this material. We are also grateful to Vladimir Pomjakushin for the HRPT experiment reported in reference \cite{White2012}. This work was supported by the Swiss National Science Foundation (Grant No. 200021-219950 and 200021-153451). This research used resources at the Spallation Neutron Source, a DOE Office of Science User Facility operated by the Oak Ridge National Laboratory. 
\end{acknowledgments}


\bibliography{main}

\end{document}


\preprint{APS/123-QED}

\title{Supplemental Material for: Coupled sawtooth chain exchange network in olivine Mn$_2$GeO$_4$}

\author{Vincent C. Morano}
\email{vincent.morano@psi.ch}
\affiliation{PSI Center for Neutron and Muon Sciences, Forschungsstrasse 111, 5232 Villigen, PSI, Switzerland}
\author{Zeno Maesen}
\affiliation{PSI Center for Neutron and Muon Sciences, Forschungsstrasse 111, 5232 Villigen, PSI, Switzerland}
\author{Stanislav Nikitin}
\affiliation{PSI Center for Neutron and Muon Sciences, Forschungsstrasse 111, 5232 Villigen, PSI, Switzerland}
\author{Jonathan S. White}
\affiliation{PSI Center for Neutron and Muon Sciences, Forschungsstrasse 111, 5232 Villigen, PSI, Switzerland}
\author{Takashi Honda}
\affiliation{Institute of Materials Structure Science, High Energy Accelerator Research Organization (KEK), Tokai, Ibaraki 319-1106, Japan}
\author{Tsuyoshi Kimura}
\affiliation{Department of Applied Physics, University of Tokyo, Bunkyo-ku, Tokyo 113-8656, Japan}
\author{Michel Kenzelmann}
\affiliation{PSI Center for Neutron and Muon Sciences, Forschungsstrasse 111, 5232 Villigen, PSI, Switzerland}
\author{Daniel Pajerowski}
\affiliation{Spallation Neutron Source, Oak Ridge National Laboratory, Tennessee, USA}
\author{Oksana Zaharko}%
\affiliation{PSI Center for Neutron and Muon Sciences, Forschungsstrasse 111, 5232 Villigen, PSI, Switzerland}

\date{\today}

\maketitle


\section{Experimental Details}

\begin{figure*}
\includegraphics[width=0.8\textwidth]{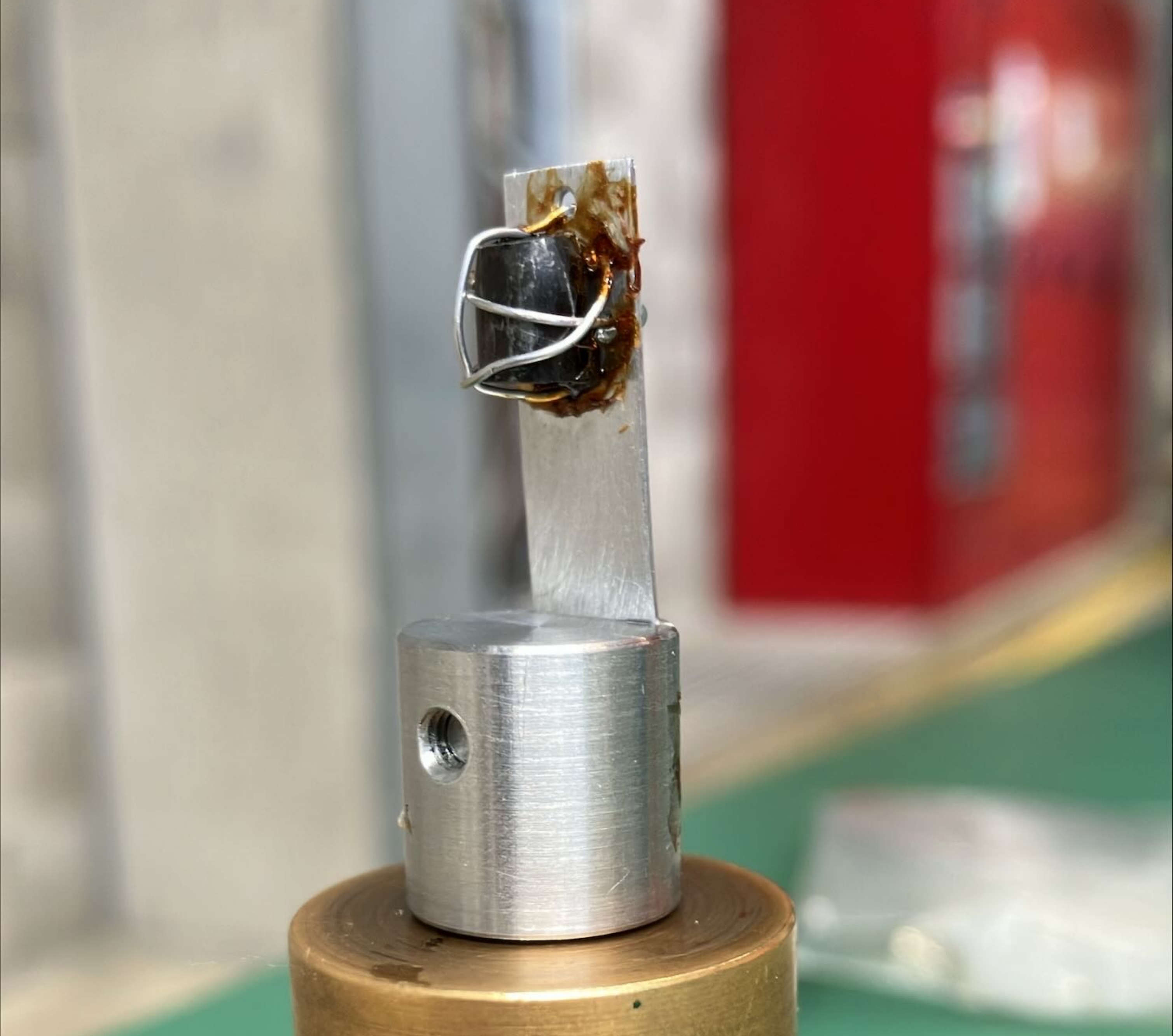}
\caption{\label{fig:SampleHolder} $0.47$ gram Mn$_2$GeO$_4$ single crystal aligned in the $(0KL)$ horizontal scattering plane.}
\end{figure*}

The $m = 0.47$ gram Mn$_2$GeO$_4$ single crystal studied in this experiment was mounted on an aluminum sample holder in the $(0KL)$ scattering plane using aluminum wire and GE varnish (Fig. \ref{fig:SampleHolder}). The sample was placed in a 5 T cryomagnet with a base temperature of $T = 1.8$ K. The High Flux chopper was selected with a frequency of 300 Hz and an incident neutron energy of $\Delta E = 12$ meV. The sample was aligned in the $(0KL)$ horizontal scattering plane to the $(0\bar{2}0)$ and $(002)$ reflections. Application of field along the vertical $\mathbf{a}$-axis did little to modify the observed spin wave excitations, here we focus on the 0-field results.

\section{Fitting Procedure}

\begin{figure*}
\includegraphics[width=0.7\textwidth]{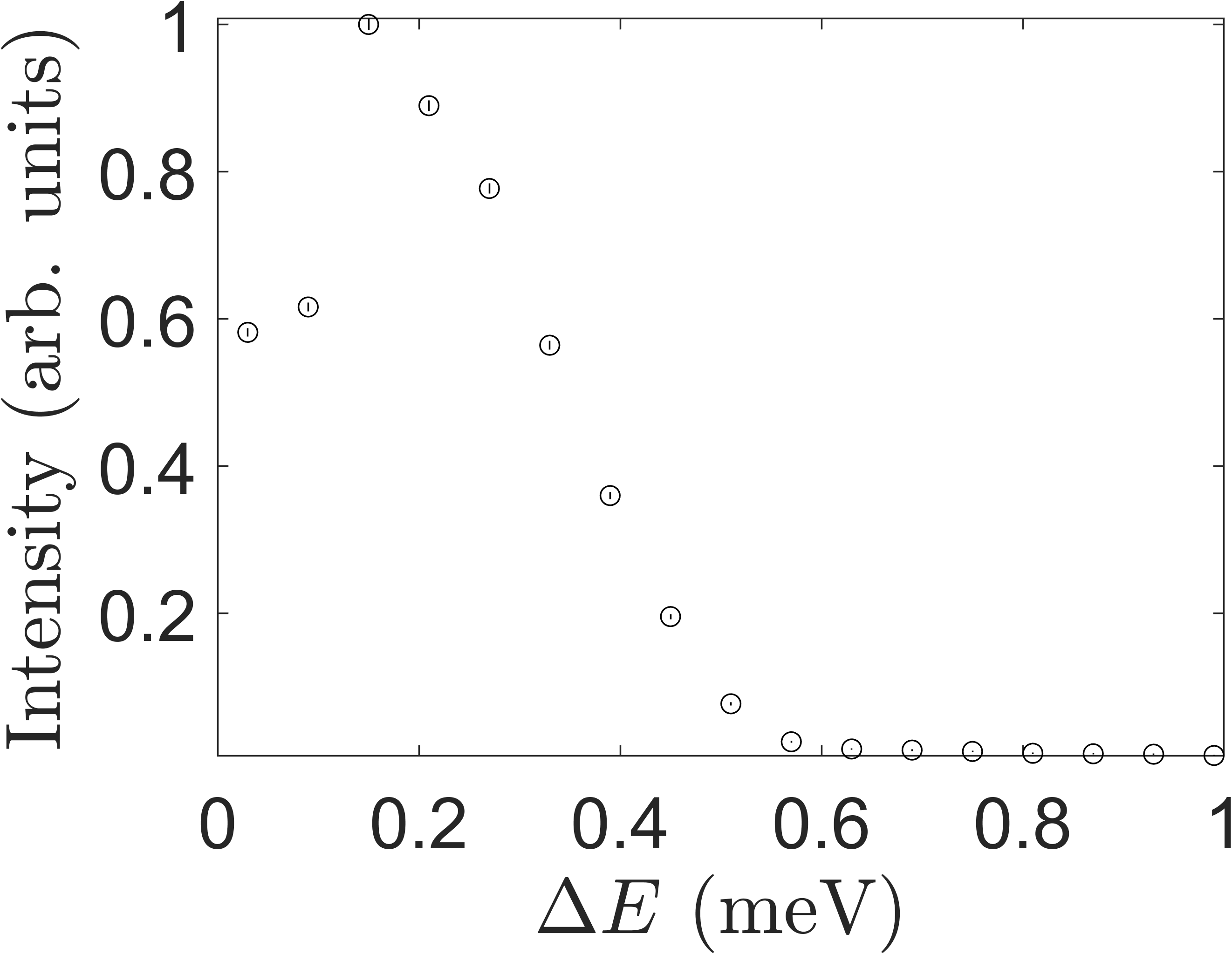}
\caption{\label{fig:Gap} Energy cut through the elastic line at $(010)$ giving a gap with upper bound $\Delta E \leq 0.2$ meV. The broadening due to the resolution at the elastic line is $\mathrm{FWHM} = 0.7$ meV. The data were integrated across 0.1 rlu bins along the $(H00)$, $(0K0)$, and $(00L)$ directions.}
\end{figure*}

Linear spin wave theory and pixel-to-pixel fits to the experimentally obtained spin wave spectrum are performed using SpinW and Horace. For a given sample rotation ($\phi$) scan at a given temperature, individual measurements are combined into a single SQW file. The SQW file is subsequently symmetrized by folding into the positive $(HKL)$ octant. The $\mathbf{Q}-\Delta E$ slices from Figures 2, 3, and \ref{fig:20K} are integrated over 0.08 \AA$^{-1}$ steps along $\mathbf{Q}$, 0.1 \AA$^{-1}$ and 0.2 \AA$^{-1}$ perpendicular to $\mathbf{Q}$ for the left and three right panels, respectively, and 0.2 meV steps in $\Delta E$. Constant-$\Delta E$ slices are binned over 0.05 \AA$^{-1}$ steps within the plotted slice, 0.1 \AA$^{-1}$ perpendicular to the slice, and 0.2 meV in $\Delta E$. In this paper, we report fits to the $T = 10$ K, $\mu_0 H$ = 0 T dataset. Ten $\mathbf{Q}- \Delta E$ slices with components spanning the 4D $\mathbf{Q}- \Delta E$ space are selected: along $(H20)$, $(H,1.5,0)$, $(0K0)$, $(0.5,K,0)$, $(0,K,0.5)$, $(00L)$, $(01L)$, $(0.5,0,L)$, $(0.5,1,L)$, and $(0,1.5,L)$. These slices are corrected for the $T = 10$ K Bose population factor then masked at the detector edges and elastic line (Fig. \ref{fig:Gap}), giving a total of 5,677,363 pixels. The fit includes an overall scale factor, a constant background term for each of the ten individual slices, and the first ten nearest-neighbor (nn) interactions for the Hamiltonian Eq. (1) discussed in the main text. The dipole-dipole anisotropy, which contains no fitted parameters, is calculated out to a threshold interatomic distance of 7 \AA ~in order to make the calculation tractable. (In Figures \ref{fig:CompareSpecSI010}b, \ref{fig:CompareSpecSI001}b we present the best-fit simulation instead using a 50 \AA ~threshold, and Figures \ref{fig:CompareSpecSI010}c, \ref{fig:CompareSpecSI001}c using the Ewald summation. The simulations are not significantly different given the instrumental resolution, supporting the 7 \AA ~threshold used in the global fitting.) The spectrum is convolved with the instrumental energy-resolution determined with the online MCViNE calculator for CNCS in our $E_i = 12$ meV, High Flux chopper configuration (https://rez.mcvine.ornl.gov/). This resolution function is expressed as a Gaussian with full width half maximum (FWHM) given by the polynomial $\textrm{FWHM} = 3.8775e-5 \omega^3 + 0.0018964 \omega^2 - 0.078275 \omega + 0.72305$ where $\omega$ is the energy transfer.

\begin{figure*}
\includegraphics[width=0.99\textwidth]{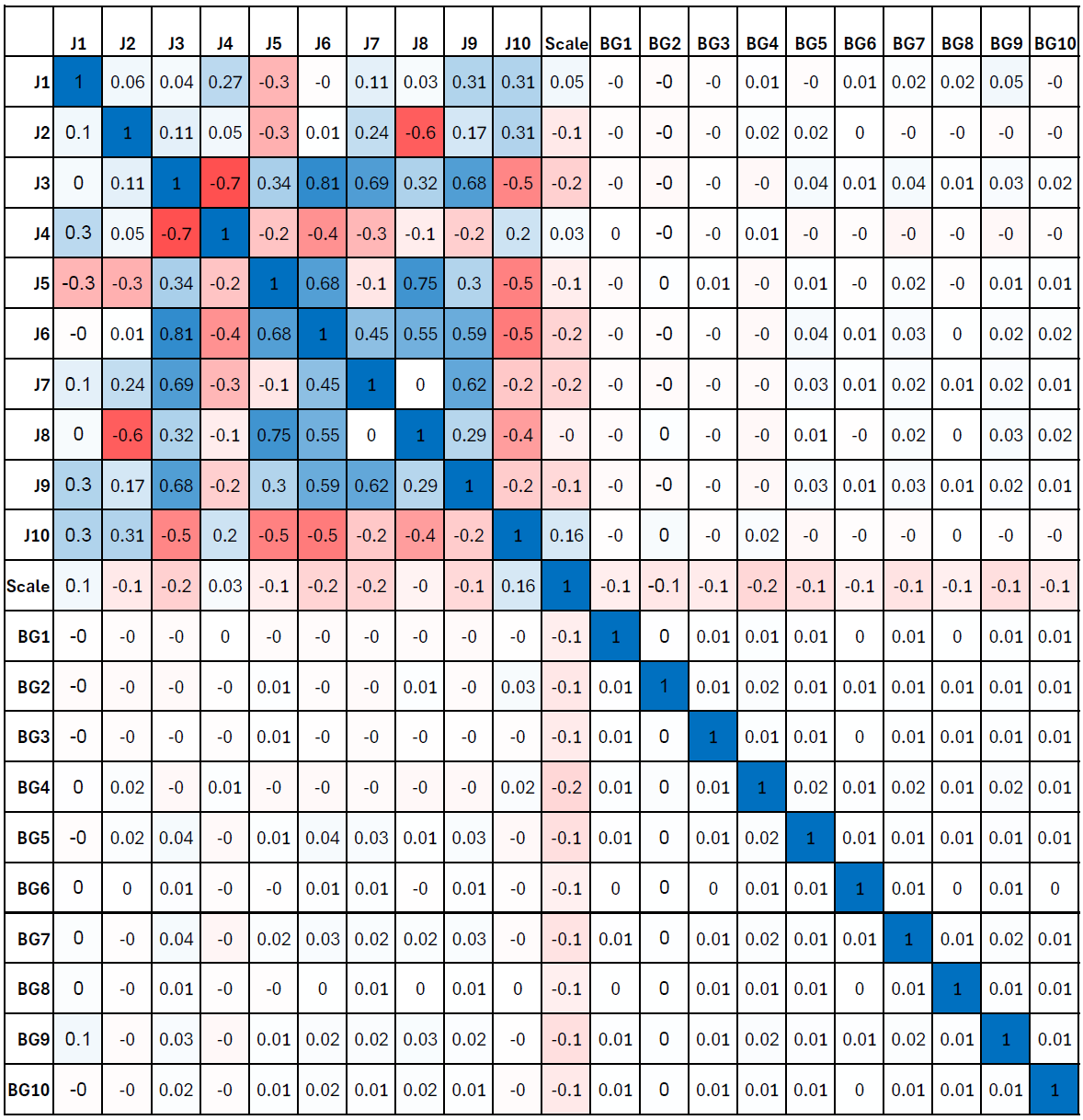}
\caption{\label{fig:covar}Covariance matrix from fitting the $T = 10$ K INS spectrum using linear spin wave theory for the Hamiltonian given by Equation 1. ``J$i$" is the $i$-th nn interaction, ``Scale" is an overall scale factor, and ``BG$j$" is a constant background term for the $j$-th $Q-\Delta E$ slice. Blue indicates parameters with positive covariance, red parameters with negative covariance, and white parameters with no covariance. Fitted values are reported in the ``Fit" column of Table \ref{tab:interactions}.}
\end{figure*}

\begin{figure*}
\includegraphics[width=0.99\textwidth]{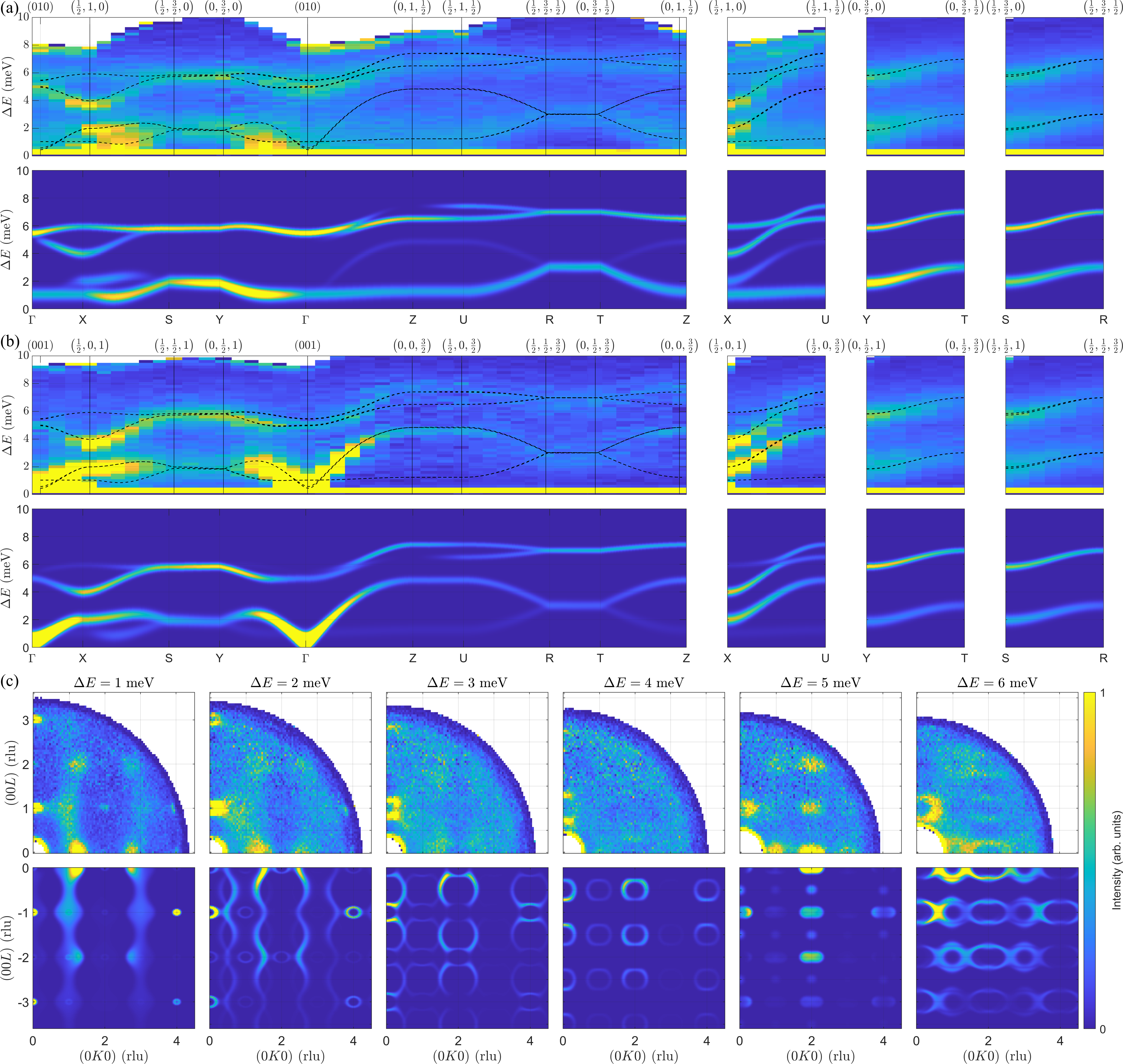}
\caption{\label{fig:20K}$T = 20$ K spectrum compared to the best-fit $T = 10$ K model. The data have been corrected for the Bose population factor at $T = 20$ K, while the simulation is performed at $T = 0$ K.}
\end{figure*}

The best-fit values are reported in Table \ref{tab:interactions} and the covariance matrix is shown in Figure \ref{fig:covar}. While the $T = 10$ K dataset is shown in the main text, Figure \ref{fig:20K} displays the $T = 20$ K dataset. There are strong correlations between the fitted parameters. Among the largest parameters, there is a large negative covariance $J_3$ and $J_4$. Physically these bonds closely resemble each other so this covariance is expected. The experiment remains sensitive to the difference between these interactions. While only Hermitian solutions are requested in the SpinW fitting routine, we find relaxing the structure in Sunny using the best-fit Hamiltonian returns imaginary eigenvalues near wavevectors of the form $\mathbf{q} = (0,0.2,0)$. These correspond to minima in the fitted dispersion, as discussed in the main text. $\mathbf{q} = (0,0.2,0)$ is near the reported incommensurability (IC) in $K$ in the multiferroic phase. Taking the Hamiltonian to be purely antiferromagnetic (AFM) and relaxing the structure in Sunny, however, gives real eigenvalues. This is true both for dipole-dipole anisotropy calculated out to a threshold distance 7 \AA, 50 \AA, and for dipole-dipole anisotropy calculated via an Ewald summation.

\section{Langevin Dynamics}

Langevin dynamics are performed in Sunny for three Hamiltonians (see the last three columns of Table \ref{tab:interactions} and Fig. \ref{fig:FETransition}).

These calculations are performed in the dipole approximation on a supercell of size $10 \times 20 \times 20$ in lattice parameters. The spins are first randomized and a Langevin integrator with damping of 0.2 and time step of 0.03 is used. The system is initially thermalized to the highest-temperature over 3000 time steps. Static pair correlations are sampled 200 times, with 100 extra time steps in between to decorrelate the samples. The instantaneous structure factor and the resulting intensities at $(100)$ and $(120)$ reflections are calculated. This process is repeated at each subsequent temperature after thermalizing over 500 time steps.

\section{Multiferroic Transition}

The AFM Hamiltonian has a minimum in the dispersion at momentum transfers of the form $\mathbf{q} = (0,0.2,0)$ up to a reciprocal lattice vector, which lies near the reported $K$-component of IC of the multiferroic phase $K_\mathrm{IC} = 0.211(2)$ but does not produce the reported $H$-component $H_\mathrm{IC} = 0.136(2)$. It was proposed \cite{White2012} that a $\mathbf{b}$-component of the nn DMI interaction becomes symmetry-allowed in the multiferroic phase and may stabilize the IC propagation vector. Including this component, with a magnitude matching $\mathbf{D}_\mathrm{1a}$, which stabilizes the $\mathbf{c}$-component of the staggered magnetization in the commensurate (C) 10 K phase, does shift the minimum of the lowest energy excitation to be IC in both $H$ and $K$. $\mathbf{D}_\mathrm{1a}$ without $\mathbf{D}_\mathrm{1b}$, however, does not stabilize the IC in $H$ nor $K$. Modestly tuning the $J$'s with both components of DMI can then condense this minimum into static long-range C+IC order. For example, uniformly decreasing $J_{3-5}$ by 0.07 meV condenses the mode, however it also significantly shifts the spin wave energies at higher  $\Delta E$. Slightly increasing $J_{1,2}$, on the other hand, condenses the low-$\Delta E$ excitation without significantly shifting the other modes.

\section{Luttinger-Tisza Calculation}

\begin{figure*}
\includegraphics[width=0.99\textwidth]{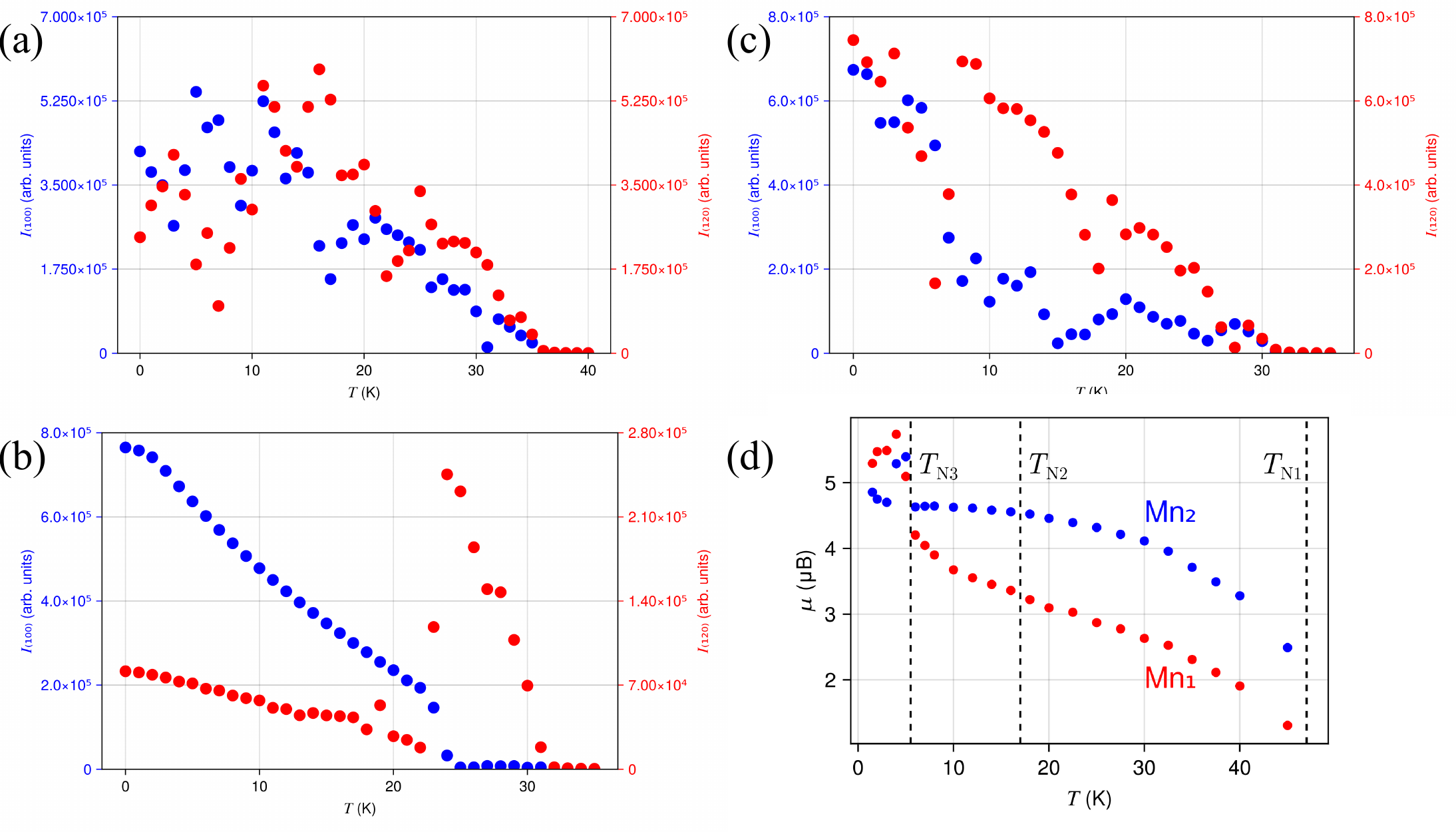}
\caption{\label{fig:Langevin}(100) and (120) reflections as a function of temperature obtained from Langevin dynamics simulations using the (a) best-fit Hamiltonian with dipole-dipole anisotropy calculated as an Ewald summation, (b) AFM Hamiltonian with dipole-dipole anisotropy calculated as an Ewald summation, and (c) AFM Hamiltonian without dipole-dipole anisotropy. While all models produce a critical onset of intensity signaling the N\'eel transition, only the pure AFM Hamiltonian with dipole-dipole anisotropy produces qualitatively corresponds to the $T_\mathrm{N2}$ spin reorientation as given in Figure 2 from reference \cite{White2012}. Note that the agreement is qualitative rather than quantitative, for instance the calculated transition temperatures are rather different from observed values. (d) $T$-dependence of the refined moment for the Mn$_1$ and Mn$_2$ magnetic sublattices refined from powder neutron diffraction data reported in the Supplemental Material of reference \cite{White2012}. Error bars are smaller than the plotted data-points.}
\end{figure*}

\begin{figure*}
\includegraphics[width=0.99\textwidth]{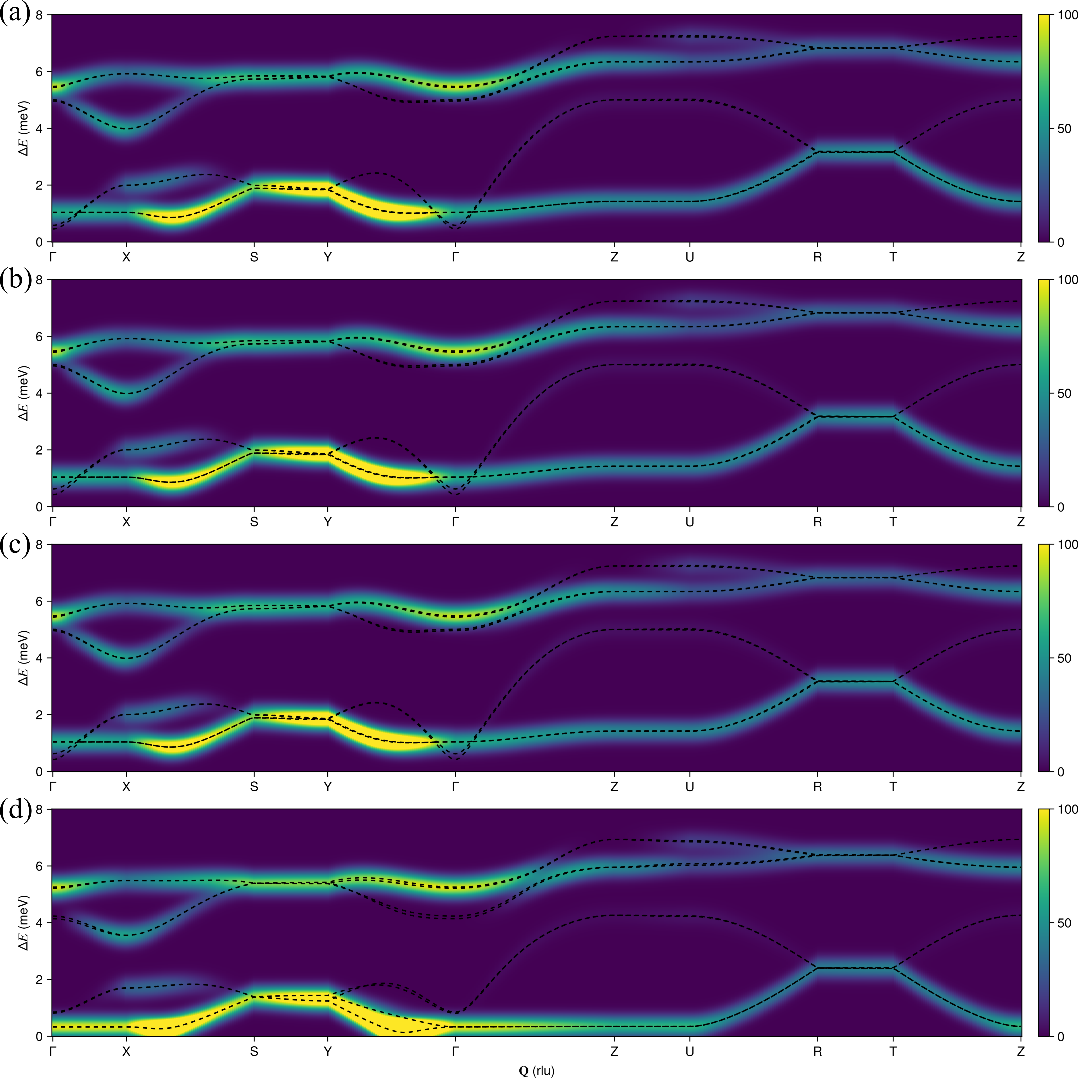}
\caption{\label{fig:CompareSpecSI010}Spectra with $\Gamma \equiv (010)$ a) using the best-fit Hamiltonian and experimentally-refined magnetic structure with a 7 \AA ~dipole-dipole threshold distance, b) the best-fit Hamiltonian and experimentally-refined magnetic structure with a 50 \AA ~dipole-dipole threshold distance, c) the best-fit Hamiltonian and experimentally-refined magnetic structure with an Ewald summation, and d) the purely AFM Hamiltonian and relaxed magnetic structure with an Ewald summation. Relaxing the structure gives imaginary eigenvalues, fixing the Hamiltonian to be purely AFM then drives modes along high-symmetry paths just above the elastic line. It also enhances the minimum along $\overline{\mathrm{Y} \Gamma}$ while removing the band crossing near $\Gamma$.}
\end{figure*}

\begin{figure*}
\includegraphics[width=0.99\textwidth]{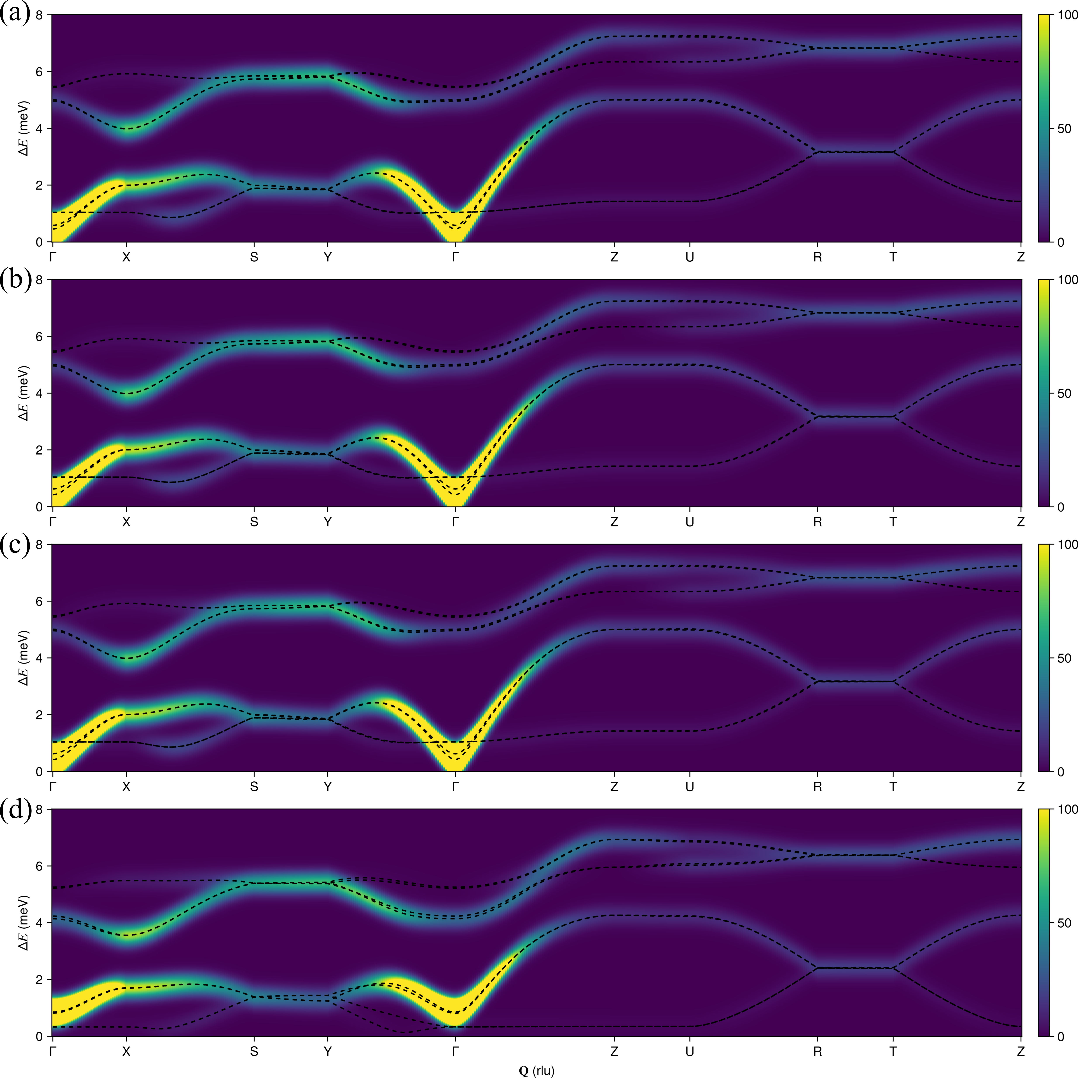}
\caption{\label{fig:CompareSpecSI001}Spectra with $\Gamma \equiv (001)$ a) using the best-fit Hamiltonian and experimentally-refined magnetic structure with a 7 \AA ~dipole-dipole threshold distance, b) the best-fit Hamiltonian and experimentally-refined magnetic structure with a 50 \AA ~dipole-dipole threshold distance, c) the best-fit Hamiltonian and experimentally-refined magnetic structure with an Ewald summation, and d) the purely AFM Hamiltonian and relaxed magnetic structure with an Ewald summation. Relaxing the structure gives imaginary eigenvalues, fixing the Hamiltonian to be purely AFM then drives modes along high-symmetry paths just above the elastic line. It also enhances the minimum along $\overline{\mathrm{Y} \Gamma}$ while removing the band crossing near $\Gamma$.}
\end{figure*}

\begin{figure*}
\includegraphics[width=0.99\textwidth]{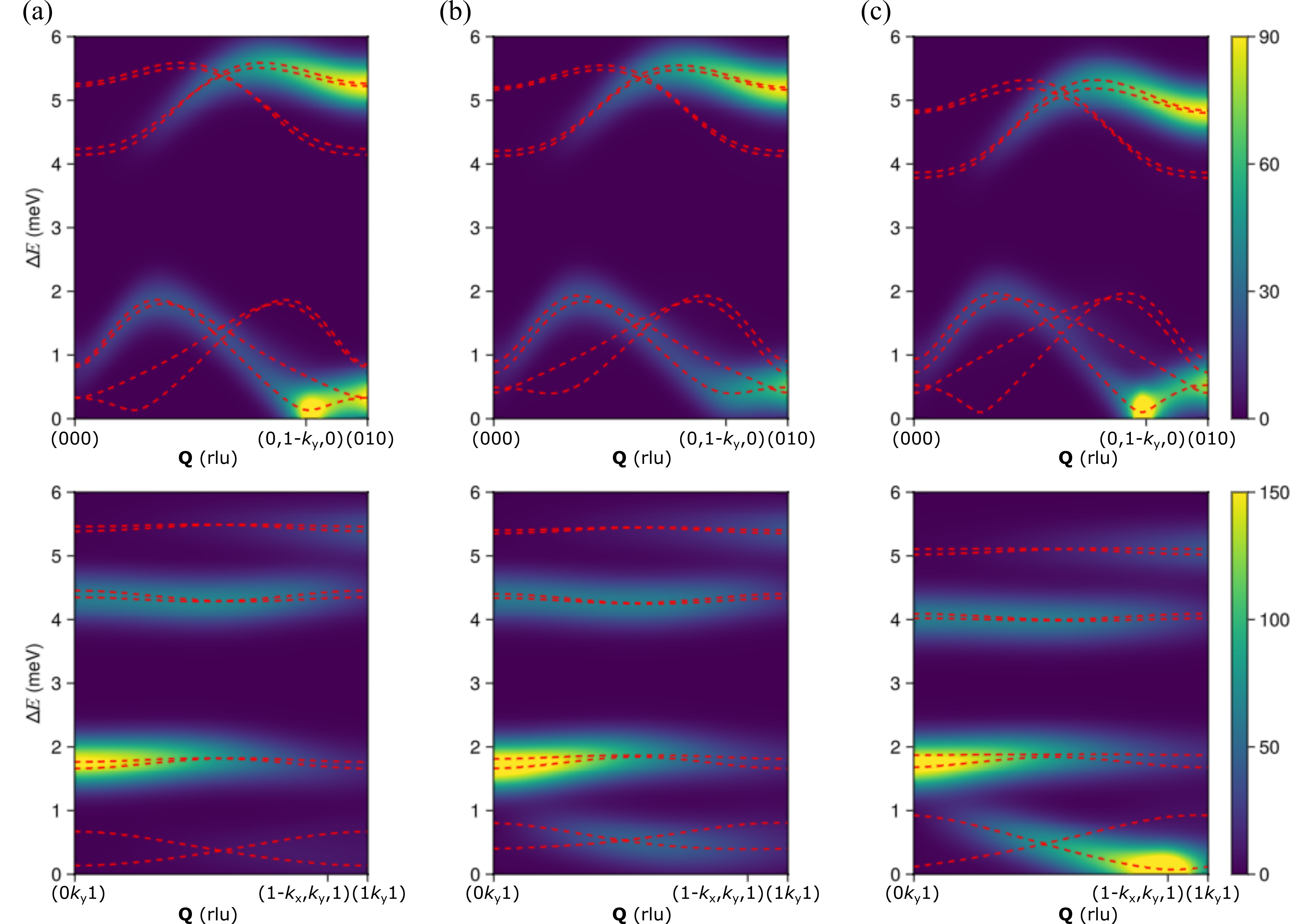}
\caption{\label{fig:FETransition}Simulated spin wave spectrum for the AFM Hamiltonian tuned to position the IC near the experimentally observed values in (a) $K$ and (b) $H, ~K$. The IC in $H$ is subtle but becomes more obvious when (c) slightly perturbing the Hamiltonian to condense these modes and stabilize the observed ferroelectric phase. See Table \ref{tab:interactions} for the Hamiltonian. Calculations performed in Sunny. $\mathbf{D}_{1\mathrm{b}}$ is included as an inhomogeneous interaction.}
\end{figure*}

Reference \cite{Santoro1966} develops a theory for the collinear to noncollinear phase transition in the olivines family based on the method of Luttinger and Tisza in the absence of dipole-dipole anisotropy. This transition occurs at $T_\mathrm{t} = 23$ K for Fe$_2$SiO$_4$ \cite{Santoro1966, Tripoliti2023} and $T_\mathrm{t} = 15$ K for Mn$_2$SiO$_4$ \cite{Santoro1966, Lottermoser1988}, near the $T_\mathrm{N2} = 17$ K transition in which the Mn$_1$ moments of Mn$_2$GeO$_4$ become increasingly noncollinear. Taking $J_1 = \gamma$, $J_2 = \gamma'$, $J_3 = \alpha''$, $J_4 = \alpha'$ and $J_5 = \alpha$, we find that inequality $2 \alpha \gamma > \gamma' (\alpha' + \alpha'')$ is satisfied. Proceeding to the analytical form of the eigenvalues and eigenvectors, we further find $2 |J_4 + J_3 - J_5|/(2*J_1 + \lambda_1) = 0.40$ and $\rho' = |J_4 + J_3 - J_2|/(2 J_1) = 0.87$, where $\lambda_1$ is the minimum eigenvalue and $\rho'$ is a critical magnetization ratio, such that $2 |J_4 + J_3 - J_5|/(2*J_1 + \lambda_1) < \rho' < 1$. This condition implies that the collinear magnetic structure is stable above a transition temperature $T_\mathrm{t}$ while the noncollinear structure is stable below. The transition temperature is determined by the relative magnetization of the two Mn sublattices $\rho = M_\textrm{I}/M_\textrm{II}$ (Figure \ref{fig:Langevin}d).

\section{Superexchange Pathways}

\begin{figure*}
\includegraphics[width=0.99\textwidth]{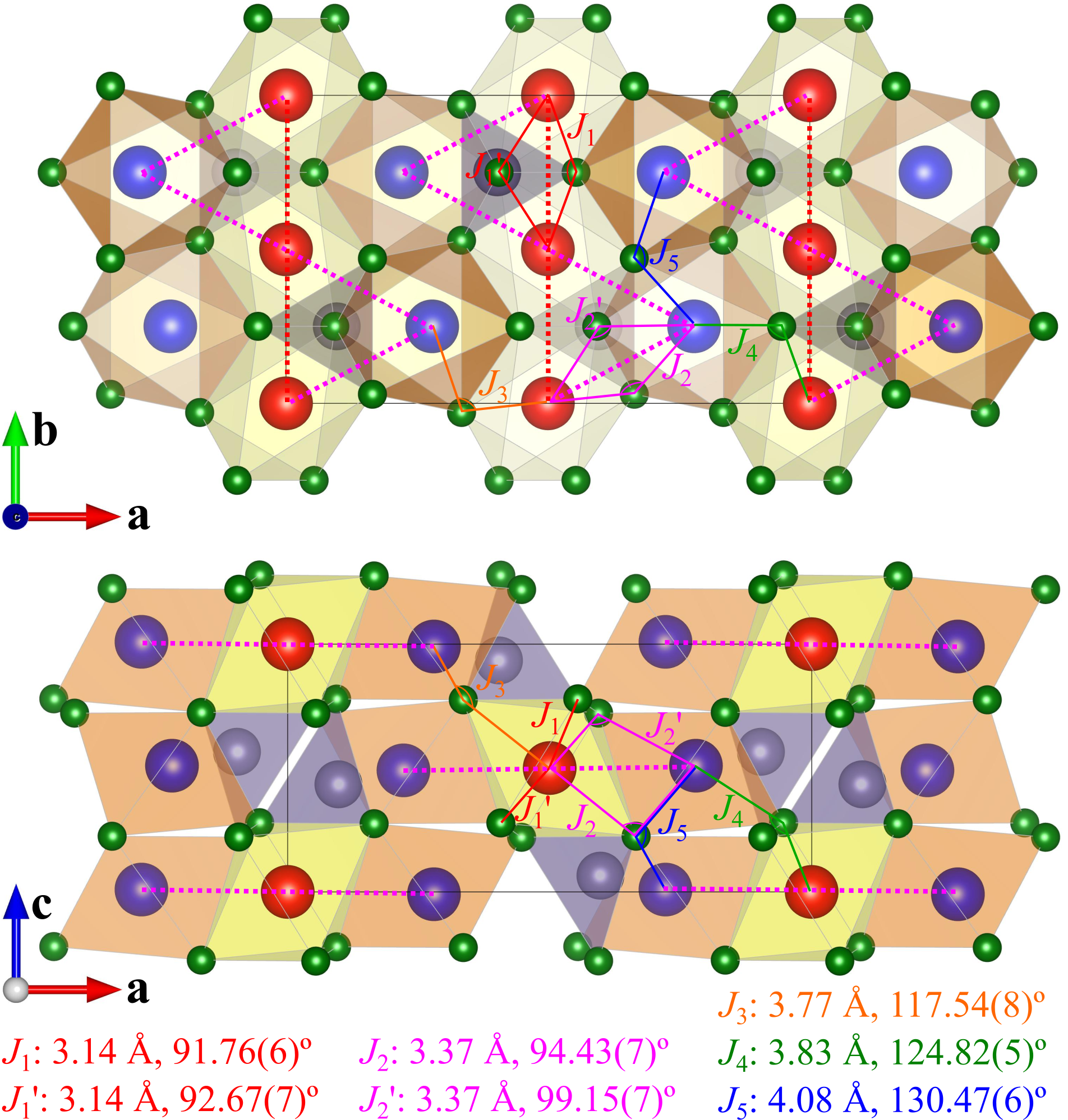}
\caption{\label{fig:Exchange} Superexchange network determined from powder neutron diffraction data obtained at $T = 7$ K in reference \cite{White2012}, including the first five nn interactions $J_{1-5}$. The nn and next nearest-neighbor interactions involve two superexchange paths, with the second path indicated by $J_1'$ and $J_2'$, respectively. Distances are the separation between Mn ions at either end of the exchange path as in Table I, while angles are those spanning the Mn-O-Mn superexchange path. $J_1-J_2$ sawtooth chains are indicated by dashed red and magenta lines. Mn$_1$ sites are red, Mn$_2$ sites are blue, all O sites are green, and Ge is gray. Image prepared with VESTA \cite{Momma2008}.}
\end{figure*}

The $T = 7$ K superexchange pathways are shown in Figure \ref{fig:Exchange}. For this 3d$^5$ system the Goodenough-Kanamori rules \cite{Goodenough1963} predict AFM interactions. While in Mn$_2$GeO$_4$ substitution of O with S \cite{Hardy1965} and Se \cite{Deiseroth2005} significantly expands the unit cell size, substitution of Ge with Si \cite{Fujii2013} to give Mn$_2$SiO$_4$ (tephroite) has little effect on the lattice parameters. Here, the principal consequence for the crystal structure is to shrink the O tetrahedra. This modifies the superexchange paths. For example, one of the two constituent $J_2$ bond angles is decreased by about 2$^\circ$ while the $J_3$, $J_4$ and $J_5$ bond angles are decreased by about 4$^\circ$, 3$^\circ$ and 2$^\circ$, respectively \cite{Yamazaki1999}.

The consequence of these modest changes in bond angle is dramatic. While Mn$_2$GeO$_4$ realizes a C+IC state below $T = 5.5$ K, breaking inversion and time-reversal symmetry to produce a multiferroic state, this transition is absent in Mn$_2$SiO$_4$. Additionally, although Mn$_2$GeO$_4$ undergoes a 17 K transition in which the net moment of each individual sawtooth chain reorients to be primarily aligned along $\mathbf{b}$, the moments in Mn$_2$SiO$_4$ cant at 15 K to gain a staggered component along $\mathbf{b}$ while the net moment of each chain remains along $\mathbf{a}$.


\bibliography{main}

\clearpage

\begin{table*}[b]
\caption{\label{tab:interactions}%
Exchange constants for the simulations displayed in Fig. \ref{fig:CompareSpecSI010}-\ref{fig:FETransition}. Columns ``IC in $K$", ``IC in $H$, $K$", and ``Condensing IC" are used for the simulations in Figure \ref{fig:FETransition}a, b, and c, respectively.}
\begin{ruledtabular}
\begin{tabular}{lccrrrr}
\textrm{Bond}&
\textrm{Distance (\AA)}&
\textrm{Atoms}&
\textrm{Fit (meV)}&
\textrm{IC in $K$ (meV)}&
\textrm{IC in $H$, $K$ (meV)}&
\textrm{Condensing IC (meV)}\\
\colrule
$J_1$ & 3.14 & Mn$_1$-Mn$_1$ & 0.478(1) & 0.478 & 0.478 & 0.498\\
$J_2$ & 3.37 & Mn$_1$-Mn$_2$ & 0.085(2) & 0.085 & 0.085 & 0.108\\
$J_3$ & 3.77 & Mn$_1$-Mn$_2$ & 0.331(5) & 0.331 & 0.331 & 0.331\\
$J_4$ & 3.83 & Mn$_1$-Mn$_2$ & 0.584(3) & 0.584 & 0.584 & 0.584\\
$J_5$ & 4.08 & Mn$_2$-Mn$_2$ & 0.303(1) & 0.303 & 0.303 & 0.303\\
$J_6$ & 5.06 & Mn$_1$-Mn$_1$ & -0.034(1) & 0 & 0 & 0\\
$J_7$ & 5.06 & Mn$_2$-Mn$_2$ & -0.017(1) & 0 & 0 & 0\\
$J_8$ & 5.58 & Mn$_1$-Mn$_2$ & -0.082(2) & 0 & 0 & 0\\
$J_9$ & 5.68 & Mn$_2$-Mn$_2$ & 0.0808(9) & 0.0808 & 0.0808 & 0.0808\\
$J_{10}$ & 5.83 & Mn$_1$-Mn$_2$ & 0.031(1) & 0.031 & 0.031 & 0.031\\
$\mathbf{D}_{1\mathrm{a}}$ & 3.14 & Mn$_1$-Mn$_2$ & 0 & 0 & 0.013 & 0.013\\
$\mathbf{D}_{1\mathrm{b}}$ & 3.14 & Mn$_1$-Mn$_2$ & 0 & 0 & 0.013 & 0.013\\
\end{tabular}
\end{ruledtabular}
\end{table*}

